\pdfoutput=1
\documentclass[acmsmall,nonacm]{acmart}

\usepackage{booktabs}
\usepackage{array}
\usepackage{amsmath}
\usepackage{placeins}

\title{Ask the Tool, Don't Guess: Agent Tool Calls Hold Their Progress, and the Serving System Should Read It}

\author{Yipeng Liu}
\email{yipeng.liu@tuna.tsinghua.edu.cn}
\affiliation{\institution{Tsinghua University}\country{China}}
\affiliation{\institution{Alibaba Cloud Computing}\country{China}}
\author{Yingqiang Zhang}
\email{yingqiang.zyq@alibaba-inc.com}
\affiliation{\institution{Zhejiang University}\country{China}}
\affiliation{\institution{Alibaba Cloud Computing}\country{China}}
\author{Feifei Li}
\email{lifeifei@alibaba-inc.com}
\affiliation{\institution{Alibaba Cloud Computing}\country{China}}
\author{Huanchen Zhang}
\email{huanchen@tsinghua.edu.cn}
\affiliation{\institution{Tsinghua University}\country{China}}

\begin{abstract}
An agentic request spends substantial wall-clock time waiting for tools, and its KV cache holds GPU memory the whole time. Serving systems decide whether that cache stays, leaves, or comes back by guessing how long the tool will run, from the tool's name, its history, a duration declared before the call, or the engine's own occupancy. We show that no estimate fixed before a call starts can know its duration, and such estimates may not even rank the calls. Meanwhile, the running tool already holds the answer, but the agent stack together with the tool silences it. We propose that tool calls report their progress explicitly while they run, and we measure what that takes. A census of four public agent corpora finds a readable signal in most tool time once it is revealed, in two strengths: a fraction of the work remaining, or an accurate signal that the end is near. A harness recovers it without changing what the agent sees, at no measurable cost to the agent's benchmark score. At the points where a KV cache decision is made, the reported progress is between several times and an order of magnitude more accurate than the best published predictors, and it stays accurate when the environment changes. Plugged into a production engine through a few small hints, it cuts the p90 time to first token (TTFT) after a tool call by 20.7\% (HBM only) and 20.8\% (HBM + DRAM) against LRU, close to an oracle. A serving system should not guess what its tools can tell it.

\end{abstract}

\begin{document}
\raggedbottom
\maketitle

\section{Introduction}
\label{sec:intro}

\begin{figure}[t]
  \centering
  \includegraphics[width=\linewidth]{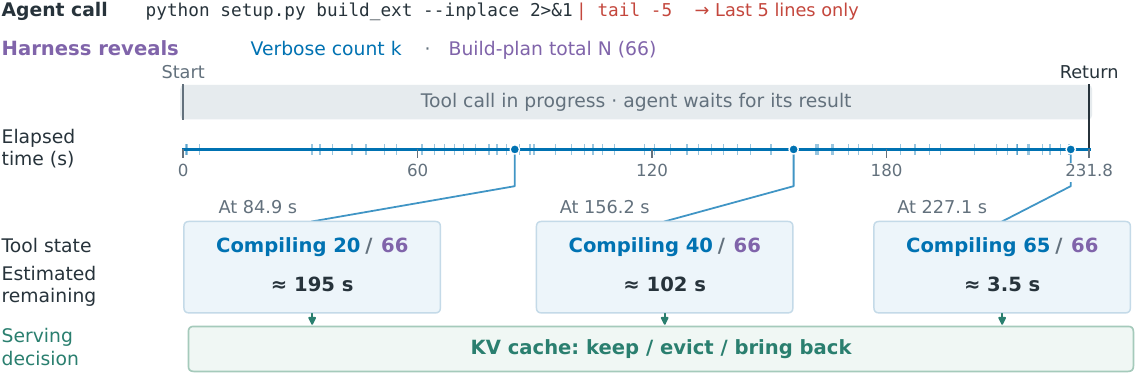}
  \caption{Progress becomes readable while the tool is still running. A real scikit-learn build (replayed) returns after 232 seconds. The agent's command pipes the output through \texttt{tail} and sees only its last five lines. The harness reads the compiler's verbose stream, which names each file as it starts, and obtains the total of sixty-six files from a dry run. Three snapshots show the count the tool holds at that moment and the remaining time it implies. These progress reports feed the decision on the KV cache side.}
  \label{fig:opening}
\end{figure}

An agentic request spends substantial wall-clock time waiting for tools \citep{speculate-tool-call}, and while it waits, its KV cache holds GPU memory that nobody is using. Between two model turns the agent runs a shell command, and the serving system must decide whether the cache stays, leaves to make room, or comes back in time for the next turn. Today it decides by guessing. It reads the tool's name, the durations of past calls, a duration declared before the call, or the occupancy of the engine \citep{continuum,cachewise,infercept,tokencake,sutradhara,sglang-rfc-24656,mooncake-rfc-2098,mori,conserve}. None of these looks at the call while it runs. The guess matters most where it is hardest. The waiting is dominated by a few long calls: the installs, builds, test suites, and generated scripts of a coding agent. A few percent of the calls hold two thirds of it. Figure~\ref{fig:opening} shows one such call from the inside, and what the agent is allowed to see of it.

The guesses fail for a reason that no better model will remove: the information is not available before the call. A long call's duration is set by the load on the machine, a neighbour in the sandbox, or the state of a remote API \citep{anthropic-infra-noise,agent-efficiency-limits,durner-s3,tail-at-scale,agentsysbench}, none of which the command line shows. Predictors built from a call's name, arguments, or history cannot even rank the long calls. History helps only when it was collected in the environment the call now runs in, and a history that is both large enough to trust \citep{continuum,cachewise} and collected under one load rarely exists (Section~\ref{sec:whyguess}).

Meanwhile, the running tool already holds the answer, and the stack silences it. The mini-SWE-agent's harness \citep{mini-swe-agent} switches progress bars off before the tool starts, and agents pipe five seconds of tool time out of six through \texttt{tail}, quiet flags, and redirects. Even with the silence lifted, the answer is not on the surface: totals are missing, counters stop halfway, and drawing the signal out without changing what the agent sees is most of the work.

We argue that the serving system should ask the tool instead of guessing. A tool call should report its progress explicitly while it runs, and revise the report as it goes, in place of a duration the system predicts implicitly before the call. What a tool can report comes in two strengths: how much work remains, or only that the end is near. Both are useful to a cache (Section~\ref{sec:census}).

We build this proposal in four steps and measure each one. We take a census of what tool calls hold across four public agent corpora and of what it takes to reveal it (Section~\ref{sec:gap}). We build a harness that recovers the signal without changing what the agent sees (Section~\ref{sec:harness}). We compare the recovered signal with four published predictors at the points where a KV cache decision is made, including under a changing environment (Section~\ref{sec:accuracy}). We add a few small primitives to a production serving engine and measure what the signal buys on GPUs (Section~\ref{sec:value}). End-to-end evaluation shows that reported tool progress cuts p90 TTFT after a tool call by 20.7\% (HBM) and 20.8\% (HBM + DRAM), close to what an oracle achieves.

The paper makes four contributions.

\begin{itemize}
  \setlength{\itemsep}{2pt}
  \item \textbf{The gap, with evidence.} We place today's timing signals on a ladder, from static caches to declared durations (Table~\ref{tab:ladder}), and show why every rung fails: estimates made before the call cannot rank the long calls, and histories are bound to the environment that produced them (Section~\ref{sec:whyguess}, Figures~\ref{fig:tail} and~\ref{fig:env}).
  \item \textbf{A census of what tools hold.} We give the first corpus-scale account of progress in agent tool calls: two strengths, five kinds, four sources. We report the share of tool time in each and what it takes to reveal it (Section~\ref{sec:census}, Table~\ref{tab:kinds}, Figures~\ref{fig:census} and~\ref{fig:shapes}).
  \item \textbf{A harness that reads it without changing what the agent sees.} We lift three layers of silence, turn output and environment into one event stream, and measure the result: accurate where the decision is made, free for the agent (Sections~\ref{sec:harness} and~\ref{sec:accuracy}, Figures~\ref{fig:harness} to~\ref{fig:trigger}, Tables~\ref{tab:triage} and~\ref{tab:arms}).
  \item \textbf{What it buys.} We add a few engine primitives, evaluate end to end on GPUs with and without a host memory tier, and bound what a lying session can gain with a session-credit mechanism that treats the tool's report as untrusted input (Section~\ref{sec:value}, Figure~\ref{fig:e2e}).
\end{itemize}

\noindent In one sentence: agent tool calls already hold their progress, a harness can read it without changing what the agent sees, and a serving system that reads it does better than one that guesses.

\section{Duration Is Decided at Run Time, and the Tool Already Holds It}
\label{sec:gap}

Picture a coding agent that has just typed \texttt{python setup.py build\_ext -{}-inplace} in a scikit-learn checkout \citep{scikit-learn} and is now waiting. Somewhere in the serving system a scheduler is asking how long this will take, but all the scheduler has is the command line. The compiler, meanwhile, is on its fortieth of sixty-six files, and it would say so, except that the harness switched the counters off and the agent piped what was left into \texttt{tail} (Figure~\ref{fig:opening}).

This section makes two points about that scene. First, no estimate made before the call starts can tell how long it will take. Second, the running tool already holds the answer: sometimes the tool prints the answer and the agent stack hides it, and more often the answer has to be drawn out. Section~\ref{sec:whyguess} gives the evidence for the first point and places today's systems on a ladder of timing signals. Section~\ref{sec:census} is a census of what the tools hold, and of what it took us to reveal it.

\subsection{Why guessing fails}
\label{sec:whyguess}

\begin{figure}[t]
  \centering
  \includegraphics[width=\linewidth]{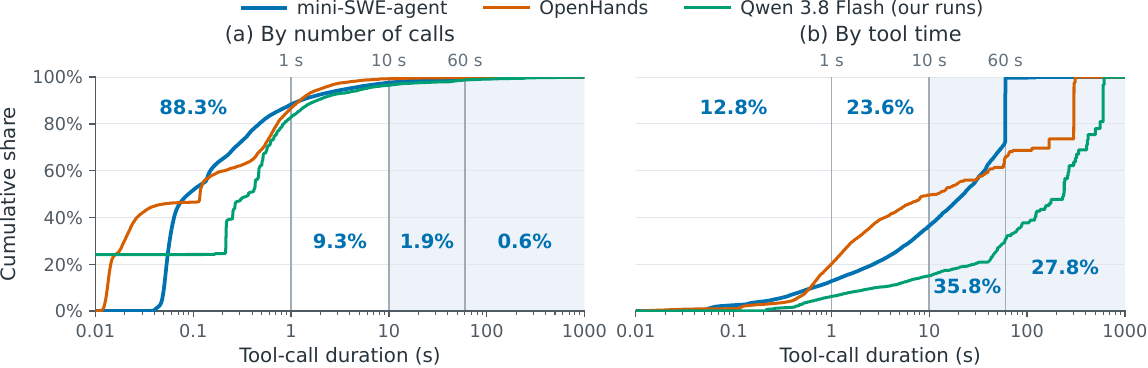}
  \caption{Long calls dominate tool time. Cumulative share of tool calls (a) by number and (b) by recorded duration, for the leaderboard runs of mini-SWE-agent \citep{mini-swe-agent,swebench-verified}, the OpenHands evaluation runs \citep{openhands}, and our own runs; annotations give the leaderboard's share in each interval.}
  \label{fig:tail}
\end{figure}

The waiting is concentrated in a few long calls. In the SWE-bench \citep{swebench,swebench-verified} leaderboard runs of mini-SWE-agent \citep{mini-swe-agent,sweagent}, fewer than three percent of tool calls hold almost two thirds of all tool time (Figure~\ref{fig:tail}): the installs, builds, test suites, and generated scripts a coding agent cannot avoid. Other traces of agent tool calls show the same concentration \citep{tracelab,mori}. For those long calls the retention decision is worth making well; for the sub-second others, almost any policy will do.

\begin{figure}[t]
  \centering
  \includegraphics[width=\linewidth]{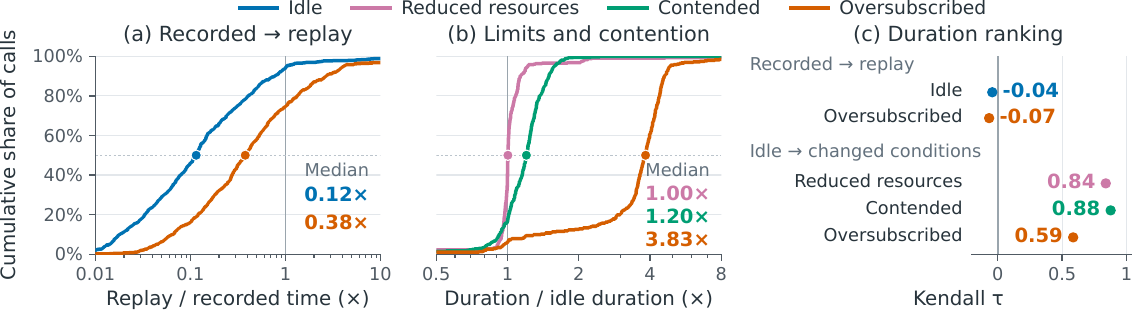}
  \caption{The same command, different environment. (a) Ratio of replayed to recorded duration for the commands of the leaderboard runs. (b) Ratio of duration under various conditions to the idle replay. Dots mark medians. (c) Rank correlation of the long calls' durations under various conditions.}
  \label{fig:env}
\end{figure}

The duration of a long call is not a property of the command itself. A call-time estimator can learn the typical duration of a class of calls. However, it cannot learn the duration of this call on this machine. We replayed the commands recorded in leaderboard runs\footnote{Throughout we read these leaderboard runs (six submissions, about three thousand trajectories), the OpenHands evaluation runs \citep{openhands}, and our own runs on a hundred random SWE-bench Verified tasks using Qwen 3.8 Flash \citep{qwen38}.} on an idle sandbox. The same command finished an order of magnitude faster (Figure~\ref{fig:env}a). Restricting the sandbox's CPU quota changed almost nothing. Adding busy neighbours, however, made most calls several times slower (Figure~\ref{fig:env}b). The order of the long calls recorded under production load has no correlation with their order on an idle machine, and when the idle machine is oversubscribed, much of its own ordering is lost as well (Figure~\ref{fig:env}c). Contention is not the only thing outside the command. Package indices, object stores, and model APIs answer at rates that follow the hour and the weekday \citep{durner-s3,iosup-cloud-variability,agent-efficiency-limits}, and their quotas follow the caller's IP \citep{dockerhub-limits}. These patterns of outside providers can hardly be turned into features of a predictor.

\begin{table}[t]
    \footnotesize
    \centering
    \setlength{\tabcolsep}{2.5pt}
    \renewcommand{\arraystretch}{1.05}
    \begin{tabular}{@{}>{\raggedright\arraybackslash}p{0.27\textwidth}>{\raggedright\arraybackslash}p{0.49\textwidth}>{\raggedright\arraybackslash}p{0.19\textwidth}@{}}
        \toprule
        What the scheduler reads   & Example systems                                                                                                                                                         & What it sees                       \\
        \midrule
        Fixed TTL                  & CachedAttention~\citep{cachedattention}, SGLang~\citep{sglang}, Anthropic~\citep{anthropic-prompt-caching}                                                              & nothing                            \\
        \addlinespace[2pt]
        Elapsed time               & InferCept~\citep{infercept}, LAMPS~\citep{lamps}                                                                                                                        & elapsed time                       \\
        \addlinespace[2pt]
        Per-tool statistics        & Continuum~\citep{continuum}                                                                                                                                             & name                               \\
        \addlinespace[2pt]
        Learned model              & CacheWise~\citep{cachewise}                                                                                                                                             & name, arguments                    \\
        \addlinespace[2pt]
        Engine-side occupancy      & MORI~\citep{mori}, ConServe~\citep{conserve}, ThunderAgent~\citep{thunderagent}, Adaptive KV Retention~\citep{adaptive-kv-retention}                                    & engine state,\newline not the call \\
        \addlinespace[2pt]
        Declared duration          & TokenCake~\citep{tokencake}, SGLang~\citep{sglang-rfc-24656}, Mooncake~\citep{mooncake-rfc-2098}, vLLM~\citep{vllm-rfc-37003}, TensorRT-LLM~\citep{trtllm-kv-retention} & fixed estimate                     \\
        \addlinespace[2pt]
        \textbf{Reported progress} & \textbf{this paper}                                                                                                                                                     & \textbf{live progress}             \\
        \bottomrule
    \end{tabular}
    \caption{The ladder of timing signals a serving system consumes during a tool call. No existing rung sees inside the call. This work adds the last rung: a progress report emitted and revised by the running tool.}
    \label{tab:ladder}
\end{table}

A history of past calls does help, but only when it was collected in the same environment the call now runs in. A history pooled across environments is an order of magnitude worse near the end of the call than the matching history (Section~\ref{sec:accuracy}), and a scheduler cannot know which history matches. Moreover, there is rarely enough of it in one place. Continuum, for one, asks for about a hundred past calls per tool before it trusts a distribution \citep{continuum}. In our corpus only a handful of tool identities reach that count. Serving systems have kept guessing anyway, and their guesses form a ladder (Table~\ref{tab:ladder}), ordered by how much of the call each source can see. Static caches keep everything for a fixed time \citep{cachedattention,sglang,agentsysbench,anthropic-prompt-caching}. Elapsed-time estimates keep a call as long again as it has already run \citep{infercept,lamps}. Per-tool statistics turn past calls into a time-to-live \citep{continuum}. Learned models cluster the arguments \citep{cachewise}. Engine-side observers watch occupancy and idleness \citep{mori,conserve,thunderagent,adaptive-kv-retention}. The newest designs let the caller declare a duration once, before the call starts \citep{tokencake,sglang-rfc-24656,mooncake-rfc-2098,vllm-rfc-37003,trtllm-kv-retention}.

None of them reads the running call: they all keep predicting its duration implicitly, from what lies around it. Every rung fixes its estimate before the call runs, or never sees inside it. This is an old lesson from another field. Database systems showed long ago that an estimator which never executes the query has no bound on its error \citep{sql-progress-trust-2005}, and that the only reliable progress indicator is the query's own count of work done \citep{sql-progress-2004,luo-progress-db}. If the estimate cannot be made before the call, can it be read during it? The same holds here: the missing rung is the one where the running tool reports how far along it is, keeps revising that report while it runs, and guides KV cache decisions on the server side. This is what this work proposes.

\subsection{What the tool holds, and what it takes to reveal it}
\label{sec:census}

The long calls are not silent, they are silenced. The infrastructure removes one layer of signal before the tool starts: every public mini-SWE-agent \citep{mini-swe-agent} run sets the environment variables that turn off pip's and tqdm's progress bars. The agent removes a second layer itself. It pipes five out of every six seconds of tool time through \texttt{\textbar{} tail}, \texttt{-q}, or a redirect to a file in our own runs. Lifting these two layers is where the work begins. Totals are missing, counters stop halfway through a run, output is buffered until the call returns, and the most informative calls print nothing at all unless asked in the right way. We therefore built the instruments to draw the signal out of each kind of call, and we counted what each instrument yields.

What a tool holds falls into five kinds. Two of them are progress. A strong signal reports how much work remains: a fraction, or a count out of a known total. A weak signal reports only that the end is near: an accurate stop signal, with no fraction worth trusting before it. Two more kinds need no progress at all, because the call is too short to matter or because the command declares its own duration. What is left is unparsable. Table~\ref{tab:kinds} gives a real command for each kind and the share of tool time it holds after our instruments. This section reports the count; Section~\ref{sec:harness} describes the instruments and shows that the agent never sees them.

\begin{table}[t]
    \centering
    \footnotesize
    \setlength{\tabcolsep}{3pt}
    \renewcommand{\arraystretch}{1.15}
    \def\kcmd#1{{\footnotesize\ttfamily\frenchspacing #1}}
    \def\kcorp#1{{\scriptsize\itshape #1}}
    \def\kband#1{{\scriptsize\color{gray}[#1]}}
    \begin{tabular}{@{}l@{\hspace{10pt}}l r r r@{}}
        \toprule
        Kind and source                      & One real command                                         & mini-SWE-agent & OpenHands     & Ours          \\
        \midrule
        \textbf{Strong signals}              &                                                          & \textbf{38.2}  & \textbf{50.6} & \textbf{48.0} \\
        \quad \textbf{A}\; in the output     & \kcmd{pytest test\_sing... -v} \kcorp{OpenHands}         & 13.4           & 32.9          & 5.7           \\
        \quad \textbf{B}\; switch or dry run & \kcmd{pytest ... -q | tail -12} \kcorp{ours}             & 18.2           & 11.5          & 31.8          \\
        \quad \textbf{C}\; a hook            & ---                                                      & 0.0            & 0.0           & 0.0           \\
        \quad \textbf{D}\; the agent's code  & \kcmd{python test\_issue.py} \kcorp{leaderboard}         & 6.5            & 6.3           & 10.4          \\
        \addlinespace
        \textbf{Weak signals}                &                                                          & \textbf{28.0}  & \textbf{8.6}  & \textbf{23.8} \\
        \quad multi-phase build              & \kcmd{make inplace} \kcorp{leaderboard}                  & 9.6            & 2.4           & 0.0           \\
        \quad download-install               & \kcmd{pip install ... | tail -5} \kcorp{leaderboard}     & 10.0           & 2.9           & 8.7           \\
        \quad count without total            & \kcmd{bin/test sympy/vector ... -q} \kcorp{ours}         & 8.4            & 3.2           & 15.0          \\
        \quad other end signal               & \kcmd{git stash \&\& pytest ...; stash pop} \kcorp{ours} & 0.0            & 0.0           & 0.1           \\
        \addlinespace
        \textbf{Negligible}                  & \kcmd{sed -i "/codeset/d" trans.py} \kcorp{leaderboard}  & \textbf{15.5}  & \textbf{29.1} & \textbf{6.2}  \\
        \addlinespace
        \textbf{Declared}                    &                                                          & \textbf{0.4}   & \textbf{0.0}  & \textbf{10.1} \\
        \quad exact                          & \kcmd{sleep 420; grep ... test.log} \kcorp{ours}         & 0.4            & 0.0           & 8.6           \\
        \quad bound                          & \kcmd{timeout 12 django runserver} \kcorp{ours}          & 0.0            & 0.0           & 1.5           \\
        \addlinespace
        \textbf{Unparsable}                  & \kcmd{python manage.py dbshell} \kcorp{OpenHands}        & \textbf{18.0}  & \textbf{11.7} & \textbf{11.9} \\
        \bottomrule
    \end{tabular}
    \caption{What a tool call holds for the serving system. Strong signals report how much work remains. Weak signals report only that the end is near. Shares are percentages of tool time after our instruments are applied.}
    \label{tab:kinds}
\end{table}

\paragraph{Strong signals.}
They come from four sources, which differ in how far the signal is from the surface. \textbf{Source A} is the tool's own output: a test runner prints a percentage column beside its dots, and pip names each package it installs. Reading them needs only the side channel. \textbf{Source B} is a switch the tool already has, at its simplest a verbose flag or an environment variable that re-enables a progress bar. It has two harder cases: some switches give a counter but no total. Our scikit-learn build names every file it compiles with \texttt{-v} but never says how many there are. Turning the counter into a fraction needs a total obtained without disturbing what the agent reads. In other cases the switch can hardly reach the tool at all. When a package is built in an isolated environment, getting a verbose flag through to the compiler would take a chain of rewrites in the build scripts. The signal is instead read from the environment, as the object files appear on disk one by one.

\textbf{Source C} is a hook the tool exposes to its host, the way language servers and agent protocols report progress to theirs \citep{lsp-progress,mcp-progress}. It is the cleanest and the rarest source. In all our sampled calls we did not find it used once. Still, it is the form the mechanism should eventually take (Section~\ref{sec:wrapper}). \textbf{Source D} is code the agent wrote itself: a script generated in the previous turn can be given a progress line as it is generated. The added line prints the script's progress in the side channel's standard format. This source calls for generation plus an equivalence check, not for a parser.

\paragraph{Weak signals.}
Their value is not in the fraction but in the last few seconds. Those seconds are what a KV cache decision needs, and revealing them took a parser of its own. Mainly two families produce this kind of signal. The first includes the multi-phase build, a \texttt{make}-style run that carries out many compilation tasks of very different sizes and then links them. A counter over the tasks is not a clock, and we show that a time extrapolated from it is often unreliable. What is usable is the announcement of the last unit and of the link step after it. We detect both, and they arrive seconds before the call returns, enough to prefetch a context back into GPU memory or to recompute it. The second family is the download, for example \texttt{pip install}. Most installs are sub-second, but the ones that stall on the network are the ones a scheduler cares about. Of the installs that stall, those that print at all announce their final phase about a second before they return. A weak signal cannot say how much is left, but in most cases its announcement of the final phase comes early enough to start a reload or a recompute in time.

\paragraph{The remaining kinds.}
They do not need progress, and we classify them before the call runs. Calls too short to matter can be recognised from the command text alone with high precision. The few that turn out long do so because of the environment, not the command, and a short timer catches them. Calls that declare their own duration give the scheduler a deadline rather than an estimate: a \texttt{sleep} is exact, and so is a \texttt{timeout} wrapping a command that would never return on its own; around any other command the timeout is an upper bound, which is what a prefetch needs. Calls that cannot be parsed are the boundary of the approach, for example a long remote API call, or a command that waits for human feedback.

\begin{figure}[t]
  \centering
  \includegraphics[width=\linewidth]{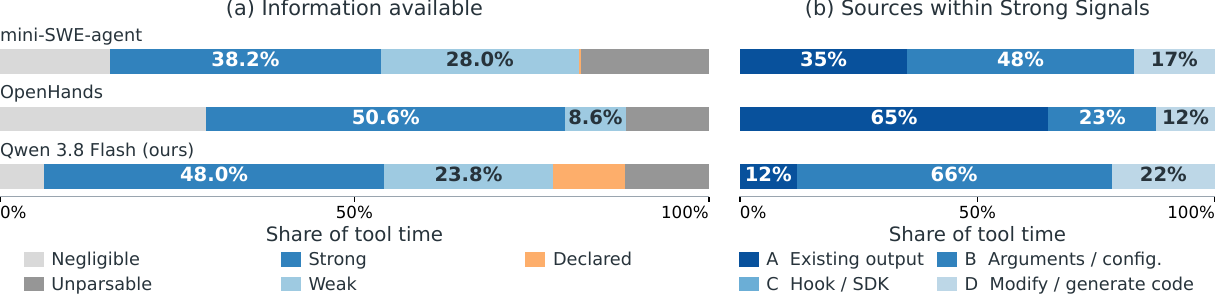}
  \caption{What tool calls hold after our instruments are applied. As share of tool time, judged after the fact with full knowledge of each call. Table~\ref{tab:arms} shows what the harness actually delivered.}
  \label{fig:census}
\end{figure}

\begin{figure}[t]
  \centering
  \includegraphics[width=\linewidth]{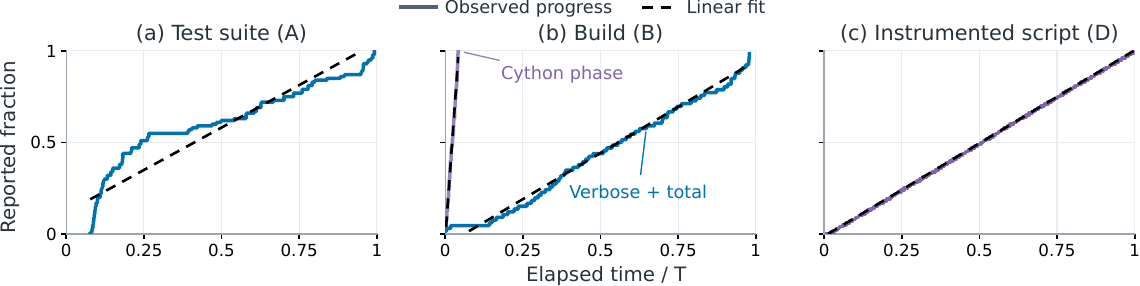}
  \caption{What the revealed signals look like. Reported fraction versus elapsed time for one long call of each kind. Dashed lines are linear fits over the observed progress stage.}
  \label{fig:shapes}
\end{figure}

The revealed signals have the shapes a scheduler needs (Figure~\ref{fig:shapes}). A test suite read from its own output is linear after a short collection phase, with the unevenness of individual tests as noise. A build read under its verbose flag, with the total from the dry run, is linear across almost the whole run. Isolated builds, read from the object files on disk, have the verbose build's shape. If read from its default output, the same build only shows the Cython phase, which does not represent the whole progress. An instrumented script is linear to the last event.

How much the agent hides depends on the model. Under the same harness and the same tasks, a different model may suppress more of its tool output and leave less of the long calls readable, but the tools hold the same things for every model. What varies is how much the agent throws away, not whether the signal exists. The tool holds it. With every instrument applied, a strong signal covers 38\% of the leaderboard's tool time, 51\% of the OpenHands runs', and 48\% of our own. A weak signal raises that to 66\%, 59\%, and 72\% (Figure~\ref{fig:census}a). The following section shows how the harness draws it out, and how an extendable execution wrapper reads it.

\section{Reading It Without Changing What the Agent Sees}
\label{sec:harness}

Back to the scikit-learn build. The compiler is on its fortieth file, and our harness has one job: get that sentence to the scheduler while the agent keeps reading exactly the bytes it would have read anyway. This section describes how. First, the rule that governs the design and the path the signal takes from the tool to the scheduler. Second, the three layers of silence, and what lifts each. Third, the wrapper that turns tool output and the tool's footprint on disk into one stream of progress events. Last, how the same design extends to tools the base model has no knowledge of.

\subsection{Three layers of silence, and what lifts each}
\label{sec:silence}

\begin{figure}[t]
    \centering
    \includegraphics[width=\linewidth]{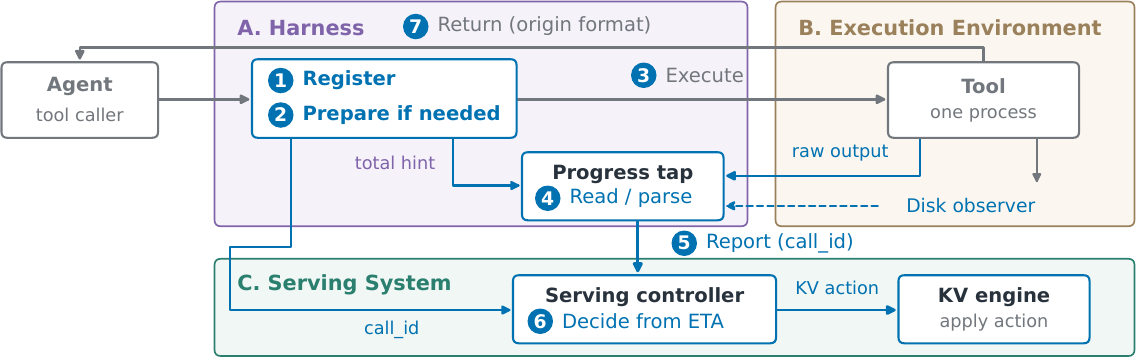}
    \caption{One execution, two consumers. The harness registers the call, prepares the command when needed, and executes the tool. While the tool runs, a tap reads its output and the files it leaves on disk, and reports its progress under the call id. Grey is the unchanged result path; blue is the state path this paper adds.}
    \label{fig:harness}
\end{figure}

One rule governs everything in this section: the agent's observation should not change. The result path to the agent is therefore untouched, and progress leaves the execution environment by a second, separate path (i.e., a side channel). Figure~\ref{fig:harness} shows the two paths of one execution: the unchanged result path to the agent, and the state path to the serving system. Before a tool call starts, the harness registers it under a call id with the serving side and prepares the command where needed: a switch, a dry run for the total, or an instrumented script. It then executes the tool. While the tool runs, an executor-side tap reads two things: the stream itself, or a side copy of it, and the files the tool leaves on disk. It reports timestamped counts, totals, and phase markers under the call id, and the serving side decides from the remaining-time estimate what to do with the request's KV cache. When the tool returns, its output and exit status reach the agent in their original formatting. Every piece lives in the harness with minimal modification to the serving engine.

The first layer of silence comes from the infrastructure and from the agent's own habits. Our harness restores the progress bars that the environment variables had turned off, and strips them again from the stream the agent reads, so the bars exist only on the side channel. A \texttt{-q} flag is reversed in the same way. Where the agent cut the output with \texttt{tail}, or sent it into a file, the harness duplicates the full stream into a side file inside the execution environment. \texttt{tail} still reads the same last lines, the file is unchanged, and so is the exit code. Section~\ref{sec:accuracy} also measures the variant in which the agent simply sees the full output. Together these rewrites put about two thirds of the suppressed tool time back on the side channel.

The second layer of silence is that many tools speak only when asked in the right way. Lifting it means asking on the agent's behalf and keeping the answer on the side channel. A verbose flag makes the scikit-learn build name every file it compiles, but not how many files there are. The harness first runs the build's own dry run to obtain the total, then adds the flag. The verbose lines go to the side channel and are removed from the stream the agent reads, so the agent's stderr is byte for byte what it was. When the flag cannot reach the compiler (e.g., in an isolated build), the wrapper watches the build directory instead. It counts the object files as they appear, with the total taken from the package's extension list. Test runners need no flag at all. They already print one dot per test as it finishes, without a newline. The wrapper parses the partial line as it grows, so the fraction arrives test by test instead of at the end.

The third layer is the agent's own code, and lifting it is a generation task rather than a parsing task. A script the agent writes usually has no progress output. The harness has the model add it at generation time: a progress line inside the loop, emitted in the side channel's format, with the requirement that the script's behaviour is otherwise unchanged. We check that requirement in two ways. We run the original and the instrumented script and compare their output and exit status, and we compare the agent's benchmark scores with and without instrumentation (Section~\ref{sec:accuracy}). The progress lines themselves never enter the agent's observation. Table~\ref{tab:instruments} in Appendix~\ref{app:harness} gives detailed examples of the instruments by layer of silence.

\subsection{One event stream from the wrapper, and how it extends}
\label{sec:wrapper}

The wrapper sits between the executor and the tool process. It passes the tool's bytes through unchanged, and on the side it produces events. It has two layers. One reads what the tool prints; the other reads what the tool leaves behind in the execution environment. Both produce the same kind of event, and both send it over the same side channel keyed by the call id.

The first layer turns a byte stream into events. Every event has the same shape: a call id, the work done, the total if known, the phase, and a timestamp. A registry of parsers recognises the counters and percentages of the common tools, a generic parser catches the rest, and a shim makes progress-bar libraries emit structured lines even when their display is disabled. A separate parser recognises the end markers of a tool's last phase. The second layer extends the byte stream. It reads the environment rather than the output. The object files of a build appear on disk one by one. The size of a download is recorded in the package index before the download starts. A test runner's collection phase leaves a count. Whatever trace a tool leaves of its work in the execution environment traces its progress, and the wrapper can read it without the tool saying a word. This is how isolated builds become as readable as verbose ones. Together, the two layers add less than one percent to the wall clock of a realistic call.

Two things make the stream trustworthy rather than merely present. When a counter passes the total it was given, the wrapper retracts the report and falls back to a count without a total, rather than clamp at one hundred percent and send false alarms. And a call with several phases is not left as several disconnected bars. Where earlier runs of the same command show how much of the call each phase takes, the wrapper folds the bars into one fraction with those shares as weights, and the call becomes a strong signal. Where the shares are unknown or unstable, the wrapper keeps only the last bar and reports its end as a stop signal, and the call stays a weak one.

What the base model can describe from pre-training already covers most of the tool calls we observed. Three paths extend the coverage to tools it has no knowledge of. The first is to teach the agent. For a given command, our harness can tell the model which switch turns progress on and supply the matching parser, and the agent then asks for progress itself. The second path belongs to tool owners. A tool can expose a hook to its host and report its own progress through it \citep{lsp-progress}. The Model Context Protocol defines a progress notification for exactly this purpose \citep{mcp-progress,mcp-tasks-sep1686}. What has been missing is a consumer on the serving side. The third path is automatic. A harness that records which tool calls hold the most time can scan them in the background, try their switches, and add the ones that work to its rules.

\section{How Good Is It, and What Does It Cost?}
\label{sec:accuracy}

Now the progress signals are there. The question a production system asks next is how good they are, and what they cost the agent. A KV cache decision has two ends. When memory runs short in the middle of a tool call, the system must decide whether this context is worth evicting. Near the end of the call it must decide when to start bringing the context back, and whether it is already too late. This section measures the signal at both ends: how much tool time can be sorted into its class before a call starts, and how the progress stream compares with four published predictors at those critical points. We also measure which degrades when the environment changes, what the agent paid, and how the picture changes across base models.

\subsection{Before the run}
\label{sec:setup}

\begin{table}[!t]
    \small
    \centering
    \setlength{\tabcolsep}{4pt}
    \renewcommand{\arraystretch}{1.1}
    \begin{tabular*}{\textwidth}{@{\extracolsep{\fill}}lrrrr@{}}
        \toprule
        Corpus & Precision (\%) & Recall (\%) & Sent short (\% of calls) & Wrongly sent, $\geq$10\,s (\% of time) \\
        \midrule
        mini-SWE-agent & 98 & 80 & 72.5 & 2  \\
        OpenHands      & 97 & 83 & 74.1 & 15 \\
        Ours           & 96 & 58 & 47.9 & 2  \\
        \bottomrule
    \end{tabular*}
    \caption{Sorting calls before they run. Precision and recall of the sorting by the harness, the share of tool calls it sent as negligible, and the share of tool time in the calls it wrongly sent.}
    \label{tab:triage}
\end{table}

\begin{figure}[t]
  \centering
  \includegraphics[width=\linewidth]{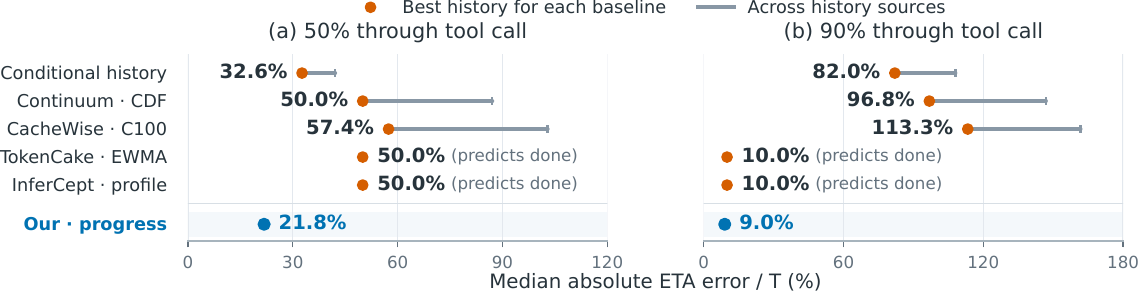}
  \caption{Remaining-time error of each estimator. The median absolute error relative to the call's duration. For each published predictor the dot is its best across history sources and the bar spans all (see Section~\ref{sec:misc}). InferCept's elapsed rule is ignored by construction; the two static predictors read ``done'' at both points.}
  \label{fig:baselines}
\end{figure}

\paragraph{Sorting calls before they run.}
Most tool time can be sent down the right path before the call starts. A lightweight classifier decides whether a call is too short to matter, declares its own duration, needs progress, or cannot be parsed. Its precision on the tool calls is high, and the calls it misses are long because of the environment rather than the command. A short timer that re-registers any call that has not returned catches them. Table~\ref{tab:triage} reports the precision and recall of that sorting per corpus, and the share of tool time in the calls it wrongly sends down the short path.

\paragraph{The competitors.}
We compare against four published predictors, each reimplemented as its paper defines it: \textbf{Continuum}'s \citep{continuum} per-tool time-to-live from the empirical distribution of past durations, \textbf{CacheWise}'s \citep{cachewise} clusters of tool names and arguments, \textbf{TokenCake}'s \citep{tokencake} running average of recent calls, and \textbf{InferCept}'s \citep{infercept,lamps} two rules, a profiled duration per tool and the elapsed rule that a call has as long left to run as it has already run. Progress signals from our instruments are converted linearly into a remaining-time estimate. We also add our own in-flight history estimator in the manner of 3Sigma's conditioning on elapsed time \citep{3sigma}.

\subsection{Accuracy at critical KV decision points}
\label{sec:critical}

\paragraph{Halfway through the call (whom to evict).}
Memory runs short for reasons outside any one call. From the call's point of view the moment is random: uniform over the call, and on average halfway through. We judge the eviction decision at the midpoint. Among the idle contexts resident in GPU memory, the serving system evicts the one whose call has the longest to run, and must not evict one whose call is about to return, because the reload would arrive too late. Both are questions about the remaining time. As a courtesy, we fix the sandbox configuration for the predictors, where their history matches the current machine. The changing-environment case comes below. Across all the methods at 50\% of each call, the progress stream's median error is about a fifth of the call's duration, against a third for the best history predictor (Figure~\ref{fig:baselines}a).

\begin{figure}[t]
  \centering
  \includegraphics[width=\linewidth]{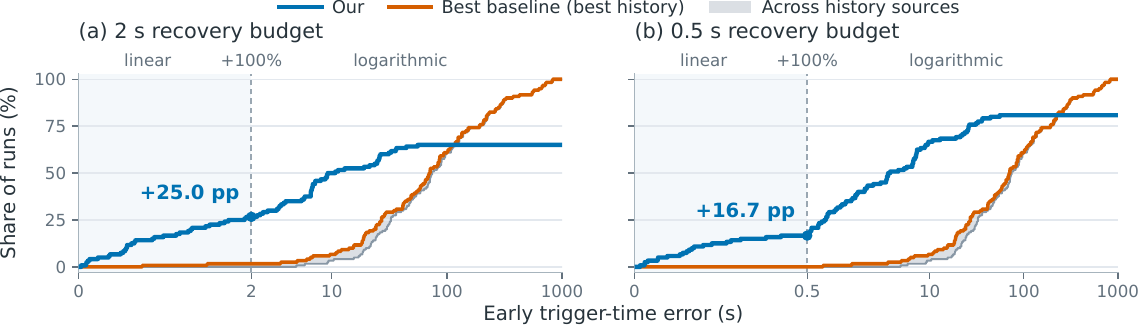}
  \caption{Timely first triggers. A run counts as timely when the estimator first reports a return no later than the recovery budget before the return, and no more than an extra budget early. The curves give the share of runs as a function of the extra lead allowed. Blue is the progress stream. Orange is the best predictor with its best history, and the band spans all history sources (Section~\ref{sec:misc}).}
  \label{fig:trigger}
\end{figure}

\paragraph{At the end of the call (when to bring it back).}
Near the end, the progress stream is far more accurate than any predictor, because the history predictors are reading the far end of a distribution that no longer describes this call. A call still running at that point has outlasted most of its class, and the history has little left to say about it. The progress stream's median error at 90\% of each call is under 10\%, against 80\% for the best history predictor (Figure~\ref{fig:baselines}b). The two static predictors that always answer ``almost done'' look as good as the stream at 90\% by construction, and they raise a false alarm at nearly every earlier checkpoint. We further quantify that error is not the whole story. What a scheduler needs is a trigger at the right moment, and that is a stricter test. The recovery budget is defined as the time it takes to get a context back, and an estimator triggers when it first reports a return within that budget. The trigger is timely if it lands just enough budget before the return. Later causes the next turn to stall, while too early puts the context in GPU memory for nothing. Figure~\ref{fig:trigger} uses two realistic budget settings. A late trigger helps nothing, so we count the runs whose first trigger is timely as a function of the extra lead time allowed. The progress stream is timely in about a quarter of the runs with a two-second budget and a sixth with a half-second budget, far better than the baselines. The appendix breaks these down by family, where uniform units do markedly better than uneven tests across tool calls.

\subsection{Robustness across settings}
\label{sec:misc}

\paragraph{Across environments.}
Guesses degrade with the environment; the progress stream does not. We replayed the same calls on the same machine under four conditions: idle; contended throughout; contended from a third of the way in; and contended until the halfway point. When we feed a predictor with matching history it may be usable, while a pool of mixed histories makes it an order of magnitude worse near the end. In contrast, the progress stream is equally accurate across the environments, because it does not depend on history. It reads the call's own rate, so a change of machine costs it only the few seconds until the new rate is observed. The matrices and the trajectories after a load change are in Appendix~\ref{app:accuracy}.

\paragraph{Across harnesses.}
Besides the production harness, we also measure the side channel alone to see what each addition brings. On our sample of SWE-bench Verified tasks, neither changes the agent's score significantly, and the side channel goes further: steps, tokens, tool time, and wall time are all unchanged as well (Table~\ref{tab:arms}). Furthermore, equivalence rests on the observations being byte-identical. A stock harness already exposes some signal. On top of that, the side channel delivers a strong signal on a fifth of the tool time, and the prompt raises that to a third, at the price of output tokens. We also tried a stronger, more intrusive variant that removes the suppression from what the agent sees. De-suppression changes the exit codes of many calls, posing a risk to the agent's behaviour. However, it does not raise the signal share at all. The longest test runs sit behind a \texttt{grep} or a \texttt{sort} in the middle of a pipeline, which no rewrite of the command's tail can undo.

\begin{table}[t]
    \footnotesize
    \centering
    \setlength{\tabcolsep}{4pt}
    \begin{tabular*}{\textwidth}{@{\extracolsep{\fill}}p{4.6cm}rrrr@{}}
        \toprule
        & stock & side channel & \textbf{+ prompt} & de-suppressed \\
        \midrule
        Strong signal (\% of tool time) & -- & $+$20.0 & $+$33.3 & $+$15.3 \\
        \quad B, switch or dry run & -- & $+$19.9 & $+$17.5 & $+$15.2 \\
        \quad D, the agent's code & -- & $+$0.1 & $+$15.8 & $+$0.1 \\
        Weak signal (\% of tool time) & -- & $+$10.2 & $+$7.6 & $+$18.8 \\
        \midrule
        & mean & \multicolumn{3}{c}{paired difference vs.\ stock} \\
        Solved (of 100 tasks) & 100 & $-$1 & $-$2 & $-$1 \\
        Steps & 35.9 & $+$0.40 & $+$2.11 & $-$0.37 \\
        Completion tokens & 14,764 & $-$458 & \textbf{$+$1,575} & $+$70 \\
        Tool time (s) & 205.1 & $-$15.2 & $-$47.2 & \textbf{$-$90.3} \\
        Wall time (s) & 495.8 & $-$34.9 & $-$24.0 & $-$82.7 \\
        Non-zero exit codes & 298 & $-$34 & $+$22 & $+$455 \\
        \bottomrule
    \end{tabular*}
    \caption{Behaviours and costs in each harness configuration. Upper rows: the share of tool time on which a strong or a weak signal reached the serving side under each configuration; the stock configuration has no tap. Lower rows: the agent's outcome and cost under stock, and how each configuration changed them, bold where the 95\,\% bootstrap interval excludes zero.}
    \label{tab:arms}
\end{table}

\paragraph{Across base models.}
The harness is model-agnostic: the same parsers, switches, and dry runs serve every model with no per-model tuning, so the signal a tool holds is the same whichever agent runs it. What changes with the model is how much of that signal survives the agent's habits. We ran the side-channel configuration on the same tasks with two open-weights models \citep{qwen38} and two closed frontier models \citep{gpt6astra,claude-fable}, each at its highest reasoning budget. They differ in how they work far more than in what they solve: one frontier model solves as many tasks as the open ones in half the steps, verifying its work throughout; the other submits after a handful of steps and rarely verifies. They differ in what they hide, too. The share of long-call time on which the harness recovers a usable stream ranges from about a third to about half across models, and follows how much output the agent suppresses, not how strong the model is.

\section{What the Signal Buys the KV Cache}
\label{sec:value}

The signal is accurate where it matters and free where it should be. We now examine its contribution to KV cache decisions. With GPU memory alone, the only decision is which idle context stays when memory runs short. With a host tier below the GPU, two more decisions appear: when to let an idle context go to the host, and when to bring it back. Both take the same signal, and the engine needs only a few hints to use it, which we added to the vLLM engine \citep{vllm} in a few hundred lines (Appendix~\ref{app:e2e}). Section~\ref{sec:setup5} sets up the replay and the two memory configurations, Section~\ref{sec:e2e} reports what the signal buys end to end, and Section~\ref{sec:trust} asks what happens when a reporter lies.

\subsection{Setup and baselines}
\label{sec:setup5}

We replay the agent sessions of Section~\ref{sec:accuracy} across all methods on a 4$\times$H100 SXM instance in two configurations: GPU memory only, and GPU memory with DRAM as an offload tier. We include two more baselines: the engine's own LRU, and an oracle that knows every exact return time. The two configurations pose different decisions. With GPU memory alone, a wrongly kept context costs someone else a recompute and a wrongly evicted one costs this call a recompute, so the decision is a pure ordering: who stays. With a host tier, an evicted context is copied to DRAM rather than dropped, so a wrong let-go costs only a reload, and the decision becomes when to let go and when to fetch back, paid for mainly in latency and DRAM traffic. Our method based on progress keeps a context while the reported remaining time is under a threshold. With a host tier it adds hysteresis, letting go only once the estimate has moved well past the threshold, and refreshes the context one lead time before the predicted return.

\subsection{End to end}
\label{sec:e2e}

\begin{figure}[t]
  \centering
  \includegraphics[width=\linewidth]{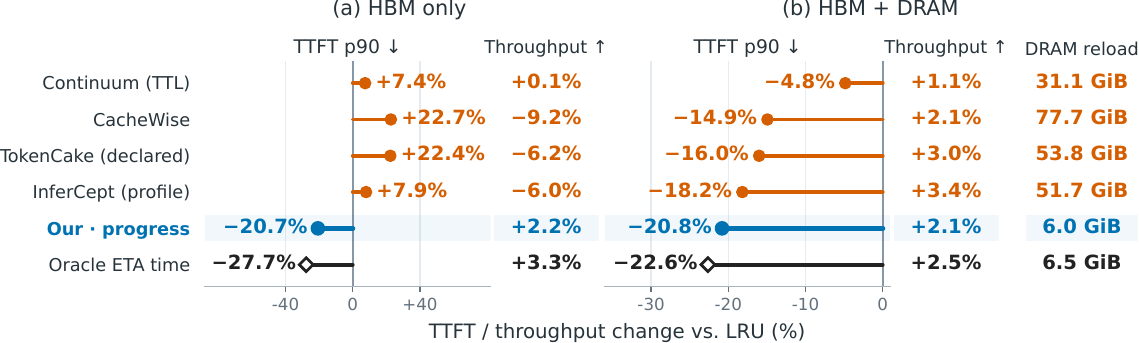}
  \caption{Post-tool TTFT and throughput in two memory configurations. (a) GPU memory only; (b) GPU memory with a host DRAM tier, with the volume each run reloads from DRAM. Blue is our method.}
  \label{fig:e2e}
\end{figure}

Figure~\ref{fig:e2e} shows the change in post-tool TTFT p90 and in throughput against LRU. With GPU memory alone, ordering by reported progress cuts the p90 TTFT after a tool call by a fifth and raises throughput slightly; the oracle cuts it by a little over a quarter. With a host tier, progress cuts the p90 by 21\% and the oracle by 23\%, and both reload a third of what LRU does. The four predictors collapse into two behaviours, and neither reads the call. Continuum's per-tool time-to-live does not expire for many calls, where it issues no hint and falls back to LRU. CacheWise, TokenCake and InferCept estimate each call from the short-dominated history of its kind, so their estimates fall below the keep threshold for most calls and they keep nearly everything. Keeping everything holds the contexts of the long calls, exactly the ones that will not return soon, in the memory that live turns need. With GPU memory alone that costs a p90 up to a quarter worse than LRU and a lower throughput. With a host tier it cuts the p90 by 15 to 18\%, but the estimates run out early, so the controller fetches contexts back from DRAM before they are needed, at eight to thirteen times our method's reload volume.

\begin{figure}[t]
  \centering
  \includegraphics[width=\linewidth]{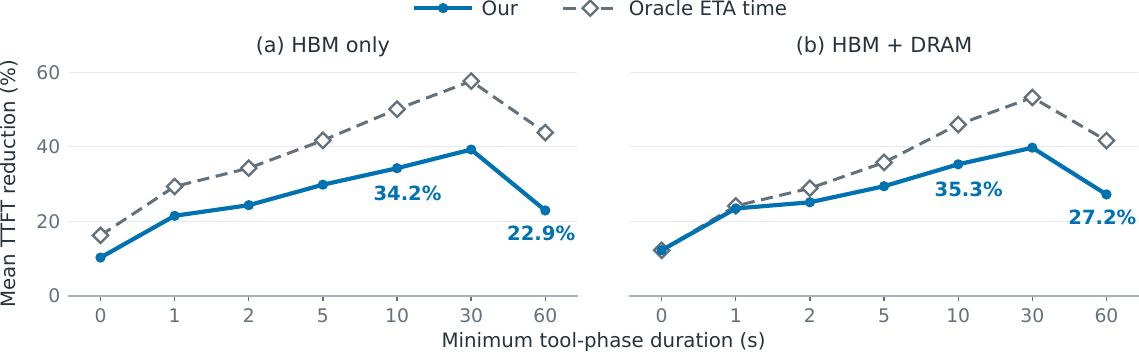}
  \caption{Mean post-tool TTFT reduction against LRU for tool calls of at least the given length, in both memory configurations. Short calls gain little; the gain grows with the length of the call, reaching 34.2\% and 35.3\% at ten seconds or more, and the oracle pulls ahead as the calls get longer.}
  \label{fig:ttftail}
\end{figure}

Figure~\ref{fig:ttftail} breaks the gain down by the length of the tool call that preceded it, as the mean post-tool TTFT reduction on the calls of at least a given length. On the whole cohort the mean gain is about a tenth, because most calls are short and nothing is evicted within a second. The gain grows with the call: about a quarter from two seconds, a third from ten, and 40\% from thirty, in both configurations. The oracle pulls ahead as the calls get longer, reaching half at ten seconds with GPU memory alone; with the host tier the two stay closer. On the few calls of a minute or more, ours falls back to a quarter while the oracle keeps most of its gain. A per-phase replay of the decisions splits the remaining distance to the oracle. Giving our method the truth only on the phases it never hears from recovers 63\% of the oracle's prize; giving it the truth only where it already has events recovers 17\%. The uncovered channel is worth about four times the covered one, so the next lever is coverage rather than precision: every step of coverage moves our method towards the oracle while the wasted refreshes fall.

\subsection{When the reporter lies}
\label{sec:trust}

A progress report is produced by content the tenant controls, so the serving system must assume that a session can lie: a script can do its heavy work last and report ninety percent done \citep{sql-progress-trust-2005,beyond-max-tokens}, and nothing in the transcript reveals it before the call returns. We therefore give no report authority on its own. Each session earns credit from reports that turned out to be right, spends it whenever the engine acts on a report, and is charged when the tool returns and the report is found wrong \citep{tsafrir-backfilling,grosof-mitzenmacher}. Two ledgers exist, one for memory kept on a report's word and one for latency risked on it, so a lie can burn only the ledger it draws on. A session with no credit left is served the way the engine serves everyone without a signal \citep{progress-bars-scheduling,wei-zhang-robustness}, so the worst a liar can do is lose the benefit, and honest sessions are unaffected. In a preliminary simulation, lying raised the memory the liars held by 6\%, and the two ledgers took 5\% of it back. In conclusion, the tool holds the signal, the harness draws it out without changing what the agent sees, an engine needs only a few hints to use it, and liars do not affect the system much.

\section{Related Work}
\label{sec:related}

Table~\ref{tab:ladder} already places the systems on one ladder by their timing signal. This section covers the three neighbours a reader will check first, then what is already known and what we add.

\paragraph{Declared durations.}
TokenCake \citep{tokencake} and the recent engine proposals that carry a field such as an expected tool duration \citep{sglang-rfc-24656,mooncake-rfc-2098,vllm-rfc-37003,trtllm-kv-retention} already let the caller state how long a tool will run, and time offload and upload against it. Every such value is fixed before the tool starts and can be corrected only when it returns. We keep that machinery and replace the constant with a stream the tool emits while it runs. The difference is measured rather than argued.

\paragraph{Predicted durations.}
Continuum \citep{continuum}, CacheWise \citep{cachewise}, and InferCept \citep{infercept} predict a tool's duration from its type, its arguments, or its history. CacheWise and InferCept measured how far such prediction sits from an oracle, on sandboxes of one fixed size. They also need a history that is large and well matched. Continuum asks for about a hundred past calls per tool before it trusts its distribution, and a history collected under one load is of little use under another. Our claim is about the decisions that need a remaining time, and about the calls whose duration is set by the environment, which no call-time feature can see.

\paragraph{Engine-side observation.}
MORI \citep{mori} and ConServe \citep{conserve} act on what the engine can see, occupancy and idleness, and ThunderAgent on phase markers the harness declares \citep{thunderagent}. None of them takes anything from the tool itself. Occupancy cannot tell a call that has just started from one that is almost finished. Our signal is a count of work already done, read from the tool itself.

\smallskip
\noindent\textbf{What is already known.}
\begin{itemize}
  \setlength\itemsep{0pt}
  \item Tools can express information to the serving system \citep{conveyor,tip}.
  \item A channel for explicit tool progress exists in the agent protocol \citep{mcp-progress,dynamo-agent-tracing}.
  \item KV caches accept hints \citep{trtllm-kv-retention,vllm-rfc-37003,sglang-rfc-36224}.
  \item Tool durations are heavy-tailed \citep{tracelab,copilot-traces}, and an estimator has no error bound \citep{sql-progress-trust-2005}.
  \item Per-client accounting of KV cache exists \citep{vtc,dlpm,prefixshield,justitia}.
\end{itemize}

\smallskip
\noindent\textbf{What we add.}
\begin{itemize}
  \setlength\itemsep{0pt}
  \item The idea of reporting progress explicitly instead of predicting it implicitly, carried through to the first serving-side consumer of an in-flight, revisable progress report.
  \item The first census of how much progress agent tool calls hold and what it takes to reveal it.
  \item The treatment of a tool-originated timing signal as untrusted input.
\end{itemize}

\section{Conclusion}
\label{sec:conclusion}

The calls that make a serving system wait are the ones whose duration cannot be known before they start, and they are also the ones that know it best. This paper measured both halves of that sentence. Call-time estimates lose the order of the long calls, and histories are bound to the environment that produced them. Meanwhile, the tools already print or can be made to reveal how far along they are, in two strengths, and a harness can read them without changing what the agent sees. Read at the moments when a KV cache decision is made, the signal is more accurate than any published predictor, and it stays accurate when the machine gets busy. A production engine then turns it into latency and throughput. What limits the gain today is coverage, and we call on tools to expose their own state for the cache to read. A serving system should not guess what its tools can tell it.

\section*{Limitations}
\label{sec:limitations}

Three boundaries of this work point in the same direction.

\paragraph{Some tool calls hold no information.}
A remote API that answers only at the end, a command that waits for a person, and a download stalled before its first byte hold nothing a harness can read, and a harness made of such calls is the boundary of our census. What would move it is on the tool side: streaming and chunked responses from the services, and a progress notification from the tools that already know their own state \citep{mcp-progress,luo-progress-compile,luo-progress-ml}.

\paragraph{Some information cannot be parsed.}
Our wrapper reads what the common tools print and what they leave behind in the execution environment, but a tool with an output format we have not met, a build system that hides its units, or a script that reports in its own words is opaque to it until a parser exists. The three extension paths of Section~\ref{sec:harness} reduce this set; they do not close it.

\paragraph{Drawing the signal out may change the tool.}
A switch, a dry run, or an instrumented script is a change to how the tool runs, even when the agent's observation is unchanged. We measured the cases we used and found them within noise, but every new instrument has to be measured again, and a tool that behaves differently under its verbose flag would break that invariant.

Each boundary is a place where the tool could help. A tool that exposed its progress, in a form it chose, would need no switch, no dry run, and no parser; the harness would only relay it. That is the direction we see: tools that expose their own state, and agent systems built to be friendly to the caches that wait for them.

\section*{Ethics Statement}
\label{sec:ethics}

Three consequences of this design should be stated. A tool's progress report is produced by content the tenant controls, so a session can lie; the credit mechanism makes lying expensive rather than impossible. A stream of progress reports reveals the structure of a tenant's workload to the serving system \citep{early-bird-timing,prefixshield}, and a deployment should treat that side channel as sensitive as the tool output itself. And when the serving system misjudges a call, the cost falls on the session that waits or on the sessions evicted to make room; the credit ledgers decide who pays, and that allocation is a policy choice, not a technical necessity. Our traces come from public benchmarks and from runs of open-weights and commercial models on them; no user data is involved.

\clearpage
\bibliographystyle{ACM-Reference-Format}
\bibliography{references}

\clearpage
\appendix

\section{Census details}
\label{app:census}

The census samples calls from four public corpora \citep{swebench,mini-swe-agent,openhands,terminalbench,exgentic-traces}, stratified by source, scaffold, command class, and duration, and labels them with two frontier language models \citep{gpt6astra,qwen38}; a third model arbitrates their disagreements, and a human-labelled subset checks them.

\begin{figure}[t]
  \centering
  \includegraphics[width=\linewidth]{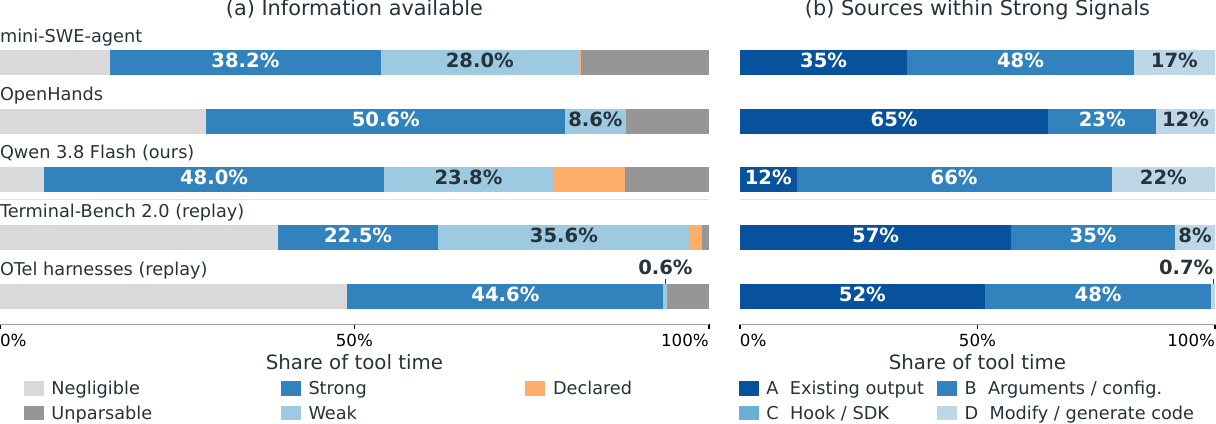}
  \caption{The census by corpus. (a) The five kinds of Table~\ref{tab:kinds} as shares of tool time. (b) Sources A--D within the strong kind, each bar normalized by its own strong time. The rows marked replay take their durations from a re-execution of the labelled calls in our sandbox environment.}
  \label{fig:appcensus}
\end{figure}

Figure~\ref{fig:appcensus} repeats the census of Figure~\ref{fig:census} for each corpus. The two SWE-bench corpora record their durations; Terminal-Bench \citep{terminalbench} and the OpenTelemetry traces \citep{exgentic-traces} do not, so their shares are of the time measured in a replay of the labelled calls. The corpora differ in what their tool time is made of. The mini-SWE-agent runs are the most mixed, with sizeable strong, weak and unparsable shares, and nearly half of their strong signal sits behind a switch (B), because the scaffold runs its calls silenced. OpenHands spends more of its time on negligible calls, views and edits, and most of its strong signal is printed by the tools as they are (A). Terminal-Bench holds the largest weak share of any corpus and the smallest strong share. The OpenTelemetry harnesses sit at the other end: half of their time is negligible and almost all of the rest is strong, split evenly between A and B. Source C is empty everywhere, and D, the agent's own scripts, ranges from a fifth of the strong signal in our own runs to almost nothing in the OpenTelemetry traces.

\section{Harness details}
\label{app:harness}

Table~\ref{tab:instruments} lists the instruments by layer of silence: what hides the signal, what the harness does about it, what the side channel receives, and what kind of signal results.

\begin{table}[t]
    \small
    \centering
    \setlength{\tabcolsep}{3pt}
    \renewcommand{\arraystretch}{1.2}
    \begin{tabular}{@{}>{\raggedright\arraybackslash}p{0.18\textwidth}>{\raggedright\arraybackslash}p{0.31\textwidth}>{\raggedright\arraybackslash}p{0.27\textwidth}>{\raggedright\arraybackslash}p{0.19\textwidth}@{}}
        \toprule
        What silences it                                         & What the harness does                                                             & What the side channel gets                                                 & Verdict                                       \\
        \midrule
        \multicolumn{4}{@{}l}{\emph{Layer 1: the infrastructure and the agent's habits}}                                                                                                                                                                                          \\
        Bars turned off by environment variables                 & Re-enable the bars; strip them from the agent's stream                            & The bar's fraction and total                                               & Weak for pip in a pipe                        \\
        Trailing \texttt{tail} or \texttt{head}, \texttt{> file} & Copy the full stream to a side file before the filter                             & The full stream; not what a mid-pipe \texttt{grep} hides                   & Transport, not a signal                       \\
        \midrule
        \multicolumn{4}{@{}l}{\emph{Layer 2: tools that speak only when asked}}                                                                                                                                                                                                   \\
        setuptools build (\texttt{build\_ext})                   & Dry run for the unit count $N$; verbose switch on, its lines hidden               & $k$ of $N$ files; last-unit marker 2 s before the exit                     & Strong                                        \\
        Isolated build (\texttt{pip install -e .})               & Count object files as they appear; $N$ from the extension list                    & $k$ of $N$ objects                                                         & Strong                                        \\
        pytest with \texttt{-q}                                  & Parse the dot line as it grows; $N$ from \texttt{-{}-collect-only}; the flag kept & $k$ of $N$ tests; \texttt{[100\%]} 0.2 s before the exit                   & Strong                                        \\
        Django's test runner                                     & Parse the dot line; \texttt{-v 2} on the side, folded back to dots                & $k$ tests; $N$ only where Django prints it; end marker 2 s before the exit & Weak below Django 4.1, strong above           \\
        \texttt{pip install} from PyPI                           & \texttt{-{}-progress-bar raw}; dry run for $N$ and the byte total                 & Bytes over the total; install marker $<$1\,s before the exit               & Strong on a slow network, weak on an open one \\
        \texttt{make}-style build                                & Verbose unit stream; \texttt{make -n} gives no $N$                                & Units without a total; last-unit marker                                    & Weak                                          \\
        \midrule
        \multicolumn{4}{@{}l}{\emph{Layer 3: the agent's own code}}                                                                                                                                                                                                               \\
        The agent's own script                                   & A progress line added at generation time; original and instrumented runs compared & $k$ of $N$ loop iterations                                                 & Strong                                        \\
        \bottomrule
    \end{tabular}
    \caption{A sample of instruments, by layer of silence. What hides the signal, what the harness does about it, what the side channel receives, and what kind of signal results.}
    \label{tab:instruments}
\end{table}

\section{Accuracy details}
\label{app:accuracy}

\begin{figure}[t]
  \centering
  \includegraphics[width=\linewidth]{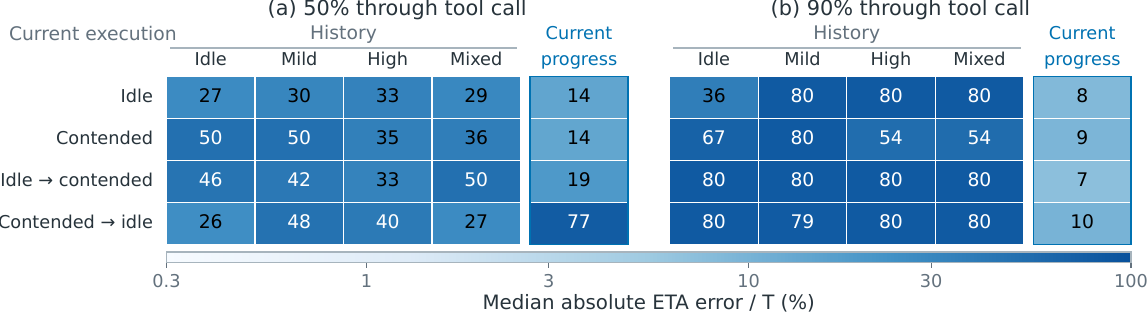}
  \caption{Median absolute ETA error at 50\,\% and 90\,\% of the call, as a share of the call's own duration, pooled over tests, builds and instrumented scripts. Rows are the load condition of the running call. Each history column shows the best method given the history source (idle, mild, high, mixed); the last column is the progress stream. Each call replayed under the four conditions.}
  \label{fig:appenvhistory}
\end{figure}

\paragraph{Error by load condition and history source.}
Figure~\ref{fig:appenvhistory} is the matrix behind the pooled numbers of Figure~\ref{fig:baselines}: the four load conditions of the running call against the four sources of history, with the progress stream in the last column. The progress stream is equally accurate on three of the four conditions. The fourth is the one where the load is removed mid-call: there the stream's rate was learned under load, so at the midpoint it overestimates the remaining time badly until it has observed the new rate, and a windowed rate shortens that lag. Contention also pushes a set of calls that were short on the idle machine past the reload threshold, where no history exists.

\begin{figure}[t]
  \centering
  \includegraphics[width=\linewidth]{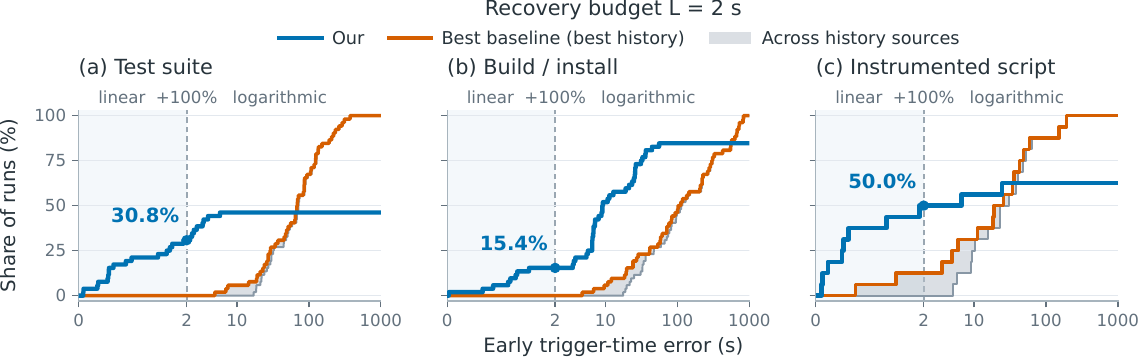}
  \caption{Timely first triggers under a two-second recovery budget, by family: the share of runs whose first threshold crossing lands within the extra lead allowed, for the progress stream (blue) and the best baseline with its best history (orange); the grey band spans the four history sources. Late and never-triggered runs count in the denominator. At the marked budget, an extra lead equal to the recovery budget, the progress stream triggers in time in 31\,\% of test runs, 15\,\% of builds and 50\,\% of scripts; the best baseline in 0, 0 and 13\,\%.}
  \label{fig:apptrigger2s}
\end{figure}

\begin{figure}[t]
  \centering
  \includegraphics[width=\linewidth]{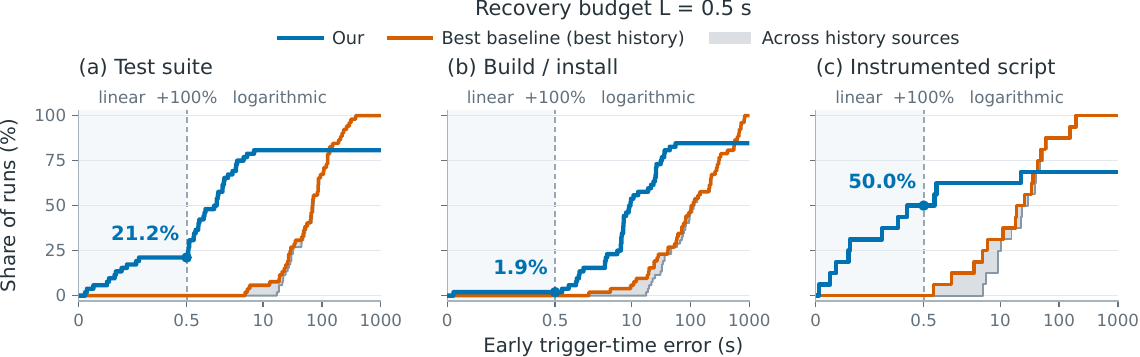}
  \caption{The same breakdown under a half-second recovery budget. At the marked budget the progress stream triggers in time in 21\,\% of test runs, 2\,\% of builds and 50\,\% of scripts; no baseline triggers in time on any run.}
  \label{fig:apptrigger0p5s}
\end{figure}

\begin{figure}[t]
  \centering
  \includegraphics[width=\linewidth]{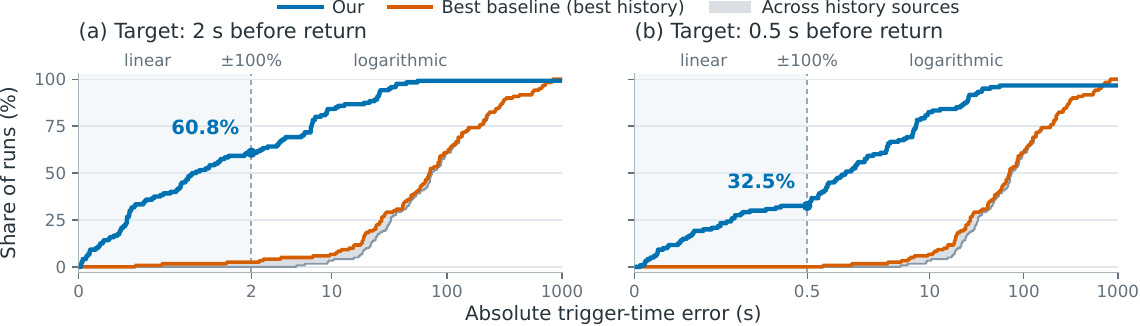}
  \caption{Absolute first-trigger time error, all families pooled: the share of runs whose first trigger lands within a tolerance of the ideal trigger time, early or late, under a two-second (a) and a half-second (b) recovery budget. Never-triggered runs count in the denominator. At a tolerance equal to the recovery budget the progress stream is within it on 61\,\% and 33\,\% of runs; the best baseline with its best history on 3\,\% and 0\,\%.}
  \label{fig:apptriggererror}
\end{figure}

\paragraph{Timely triggers by family.}
Figures~\ref{fig:apptrigger2s} and~\ref{fig:apptrigger0p5s} split the trigger curve of Figure~\ref{fig:trigger} by family under the two recovery budgets. The instrumented scripts, whose loop iterations are uniform, trigger in time in half of the runs at either budget. Test suites, whose tests are uneven, trigger in time in a fifth to a third. Builds, whose compile steps are the most uneven, rarely trigger in time on a tight budget, but do so on most runs once more lead is allowed. The best baseline with its best history triggers in time on almost no run outside the scripts. Figure~\ref{fig:apptriggererror} relaxes the one-sided count to a two-sided one: a trigger counts when it lands within a tolerance of the ideal time, early or late. Within a tolerance equal to the recovery budget, the progress stream lands close to the ideal time on well over half of the runs with two seconds of recovery and on a third with half a second, where the baselines almost never do.

\begin{figure}[t]
  \centering
  \includegraphics[width=\linewidth]{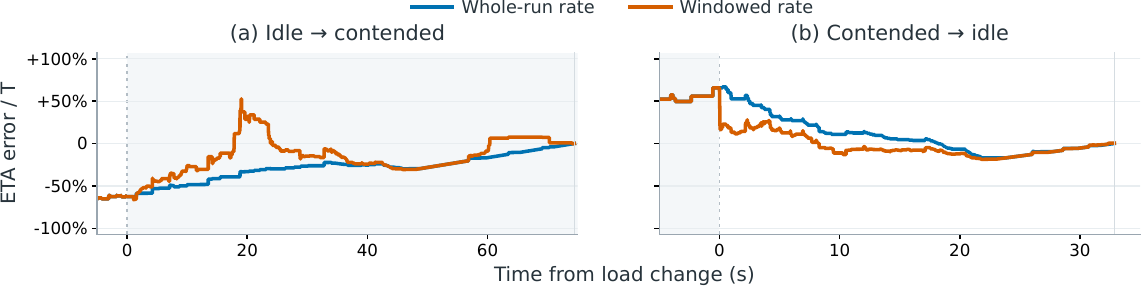}
  \caption{Online ETA error across a load change on two real replays of the same editable install, read from verbose compile output with a known total: a contending neighbour starts (a) or stops (b) at time zero, and the shaded span is the contended interval. Both estimators are reconstructed from the events available at the time; a positive error overestimates the remaining time.}
  \label{fig:appdynamics}
\end{figure}

\paragraph{Trajectories after a load change.}
Figure~\ref{fig:appdynamics} follows the online estimate across a load change on two real replays of the same editable install: a contending neighbour starts in one and stops in the other. The whole-run rate carries the old load for tens of seconds. A windowed rate recovers within about ten seconds, at the price of overshooting.

\section{End-to-end details}
\label{app:e2e}

\paragraph{The engine patch.}
The engine is vLLM \citep{vllm} with prefix caching on; the host tier is its own KV offloading to DRAM. An unmodified engine gives an outside controller no handle on an idle context: keeping one meant sending it a one-token request, a recompute that competes with real turns, and letting one go was impossible. The patch adds one route that names a context by the id of an earlier response and offers three operations: \emph{refresh} moves its resident blocks to the safe end of the eviction queue at no cost, \emph{release} moves them to the front while keeping them addressable, and \emph{query} reports how much of it is resident. The operations run between scheduler steps, skip blocks that a running request holds, and never raise into the serving loop. The scheduler, its eviction rule, and the model are untouched, and an engine that receives no hints behaves exactly as before. A hint costs one round trip of a few milliseconds, and the patch is a few hundred added lines with no existing function changed.

\paragraph{The controller.}
The controller runs beside the harness and reads each session's progress stream. While the reported remaining time is under the keep threshold, it refreshes the session's context; once the estimate has moved well past the threshold, it stops, and the context ages out under LRU or moves to the host tier; if the estimate falls back, the refreshes resume. With a host tier, the controller also refreshes the context one lead time before the predicted return and reloads it if it has gone. The oracle runs the same controller with the true remaining time; LRU never hints.

\end{document}